\documentclass[aps,prl,reprint,showpacs,superscriptaddress,groupedaddress,amsmath,amssymb]{revtex4-1}
\usepackage{graphicx}
\usepackage{dcolumn}
\usepackage{bm}
\usepackage{color}
\usepackage{ulem}
\usepackage{epsfig}
\usepackage{amsmath}
\usepackage[none]{hyphenat}

\begin{document}

\title{Principal Modes in Multimode Fibers: Exploring the Crossover from Weak to Strong Mode Coupling}

\author{Wen Xiong$^{1}$, Philipp Ambichl$^{2}$, Yaron Bromberg$^{1}$, Brandon Redding$^{1}$, Stefan Rotter$^{2}$, Hui Cao$^{1}$}
\email{hui.cao@yale.edu}
\affiliation{$^{1}$Applied Physics Department, Yale University, New Haven CT 06520, USA. \\
	$^{2}$Institute for Theoretical Physics, Vienna University of Technology (TU Wien), A-1040 Vienna, Austria.}

\date{\today}

\begin{abstract}

	We present experimental and numerical studies on principal modes in a multimode fiber with mode coupling. By applying external stress to the fiber and gradually adjusting the stress, we have realized a transition from weak to strong mode coupling, which corresponds to the transition from single scattering to multiple scattering in mode space. Our experiments show that principal modes have distinct spatial and spectral characteristic in the weak and strong mode coupling regimes. We also investigate the bandwidth of the principal modes, in particular, the dependence of the bandwidth on the delay time, and the effects of the mode-dependent loss. By analyzing the path-length distributions, we discover two distinct mechanisms that are responsible for the bandwidth of principal modes in weak and strong mode coupling regimes. Taking into account the mode-dependent loss in the fiber, our numerical results are in good agreement with our experimental observations. Our study paves the way for exploring potential applications of principal modes in communication, imaging and spectroscopy.
	 
\end{abstract}

\maketitle
\section{Introduction}
Recent advances in coherent control of light propagation in random scattering media \cite{MoskNatP12} have triggered experimental investigations of the transmission eigenchannels \cite{MoskPRL08, PopoffPRL10, ChoiPRB11, ShiPRL12, ChoiNatPhoton12, YuPRL13, PopoffPRL14, ShiPRB15, DavyNatComm15, GerardinPRL14, SarmaPRB15, SarmaPRL16}, which provide a full description of steady-state transmission of monochromatic waves. The pulsed transmission is much more complex and involves not only spatial but also temporal distortions of an input signal. As the multiple scattering creates innumerable possible paths that light can take, the temporal shape of a pulse is severely distorted and stretched. The inherent coupling between temporal and spatial degrees of freedom makes it possible to exert control over the temporal dynamics of the transmitted pulse solely by manipulating the spatial degrees of freedom of the incident wavefront. Spatiotemporal focusing has been achieved by mitigating the temporal distortion in a single spatial channel \cite{KatzNatPhoton11,McCabeNatCommun11,AulbachPRL11,AulbachOE12,ShiOL13,AndreoliSciRep15,MounaixPRL16}. A global control of pulsed transmission in all spatial channels is much more challenging, and it is not clear whether the spatial degrees of freedom are sufficient to tailor the temporal dynamics of the {\it total} transmission through turbid media.  

Multimode optical fibers (MMFs) have attracted much attention lately due to practical applications in communication \cite{RichardsonNatPhoton13, HoJLT14}, imaging \cite{ChoiPRL12, CizmarNatComm12, PapadopoulosOE13, Caravaca-AguirreOE13, BrankovNatCommun14, PloschnerNatPhoton15, LoterieOE15, AmitonovaOL16} and spectroscopy \cite{ReddingOL12, ReddingOE13, ReddingAO14, ReddingOptica14, LiewOL16}.  Intrinsic imperfections (like an inhomogeneity of the refractive index in the fiber) and external perturbations (such as  those causing a cross-section deformation) lead to coupling of the guided modes. Such coupling can be considered as optical scattering in mode space, with the effective transport mean free path $\ell$ given by the propagation distance beyond which the spatial field profile becomes uncorrelated \cite{HoJLT14, DopplerNJP14}. If the fiber length $L$ is less than $\ell$, the weak mode coupling can be described as single scattering of light from one mode to another. Once $L$ exceeds $\ell$, light is scattered back and forth among the fiber modes \cite{DopplerNJP14}. Note that the scattering occurs in mode space as light still propagates forward but in different modes. In the strong mode coupling regime, light may return to the original mode after hopping to other modes and introduce interference effects. Thus multiple scattering and wave interference become dominant. 

Light propagating through a MMF experiences spatial distortions that scramble the intensity profile. Such distortions have been effectively corrected at a single frequency by shaping the input wavefront. In fact, an arbitrary output field pattern can be generated with monochromatic light \cite{BianchiLC11, CizmarOE11}. In addition to the spatial distortions, a short pulse propagating through a MMF experiences temporal distortions. Even if a pulse is launched to a single guided mode of the MMF, the random mode coupling spreads light to other modes with different propagation constants. A selective excitation of modes with similar propagation constants results in the formation of a focused spot with minimal temporal broadening at the output of a MMF \cite{MoserOE15}. This method, however, works only when mode coupling is relatively weak, as multiple scattering spreads the input light to all modes with distinct velocities. 

To overcome modal dispersion, principal modes (PMs) were proposed for MMFs as the generalization of principal states of polarization in a single mode fiber \cite{FanOL05, ShemiraniJLT09, JuarezOE12, AntonelliOE12, NolanSPIE14, MilioneJOSAB15} and they provide an effective approach to mitigate temporal distortions in the strong mode coupling regime. A PM retains its output spatial profile to the first order of frequency variation \cite{FanOL05}. Mathematically PMs are the eigenstates of the group-delay matrix $G \equiv -iT^{-1} {dT}/{d\omega}$, where $T$ is the field transmission matrix. In the absence of backscattering in the fiber, the group delay matrix coincides with the Wigner-Smith time-delay matrix, $Q \equiv -iS^{-1} {dS}/{d\omega}$, where $S$ is the scattering matrix \cite{EisenbudThesis48,WignerPR55,SmithPR60}. Hence, PMs correspond to the Wigner-Smith time-delay eigenstates \cite{RotterPRL11}, and have well-defined delay times that are equal to the real part of the associated eigenvalues. These eigenstates provide the most suitable basis for studying and controlling temporal dynamics of total transmission through MMFs. 

In the absence of mode coupling, PMs are linearly polarized (LP) modes, i.e., the eigenmodes of a perfect fiber in the weak guiding approximation. Mode coupling entangles spatial and temporal degrees of freedom. However, the output spatial profile of a PM is decoupled from its  temporal profile. Different output spatial channels follow the same temporal trace, thus the spatial profile of the output field remains constant in time. Neglecting chromatic dispersion in the fiber, when a transform limited pulse is launched to a single PM, the output pulses in all spatial channels remain short and undistorted, even in the presence of strong mode coupling. Recent studies show that PMs in MMFs with weak and strong mode coupling have distinct spatial profiles and spectral correlation bandwidths \cite{CarpenterNatPhoton15,XiongPRL16}. An important open question that remains to be solved, however, is how the transition occurs, i.e., how PMs evolve from the weak to the strong mode coupling regime.  A physical understanding of PMs in different regimes is not only important for the fundamental comprehension of temporal dynamics of mesoscopic transport, but also relevant to applications in telecommunication and imaging.

In this paper, we experimentally study PMs in both weak and strong mode coupling regimes as well as in the transition region between them. With weak mode coupling, each PM is a mixture of a few modes with similar propagation constants, while with strong mode coupling, a PM consists of many modes. We investigate spectral correlation widths of PMs with different delay times and how mode-dependent loss affects the widths. In the weak mode coupling regime, the spectral correlation widths of PMs decrease dramatically with the increase of the delay times. However, in the strong mode coupling regime, the correlations exhibit a plateau within the short delay time range. We perform numerical simulations to further confirm and understand our experimental observations. By calculating the intensity distribution over the path-length, the finite bandwidth of PMs can be explained. Taking into account mode-dependent loss in the MMF, the numerical results show agreement with the experiment data. 

\section{Experimental measurement of principal modes}
\begin{figure}
\centering
\includegraphics[scale=0.6]{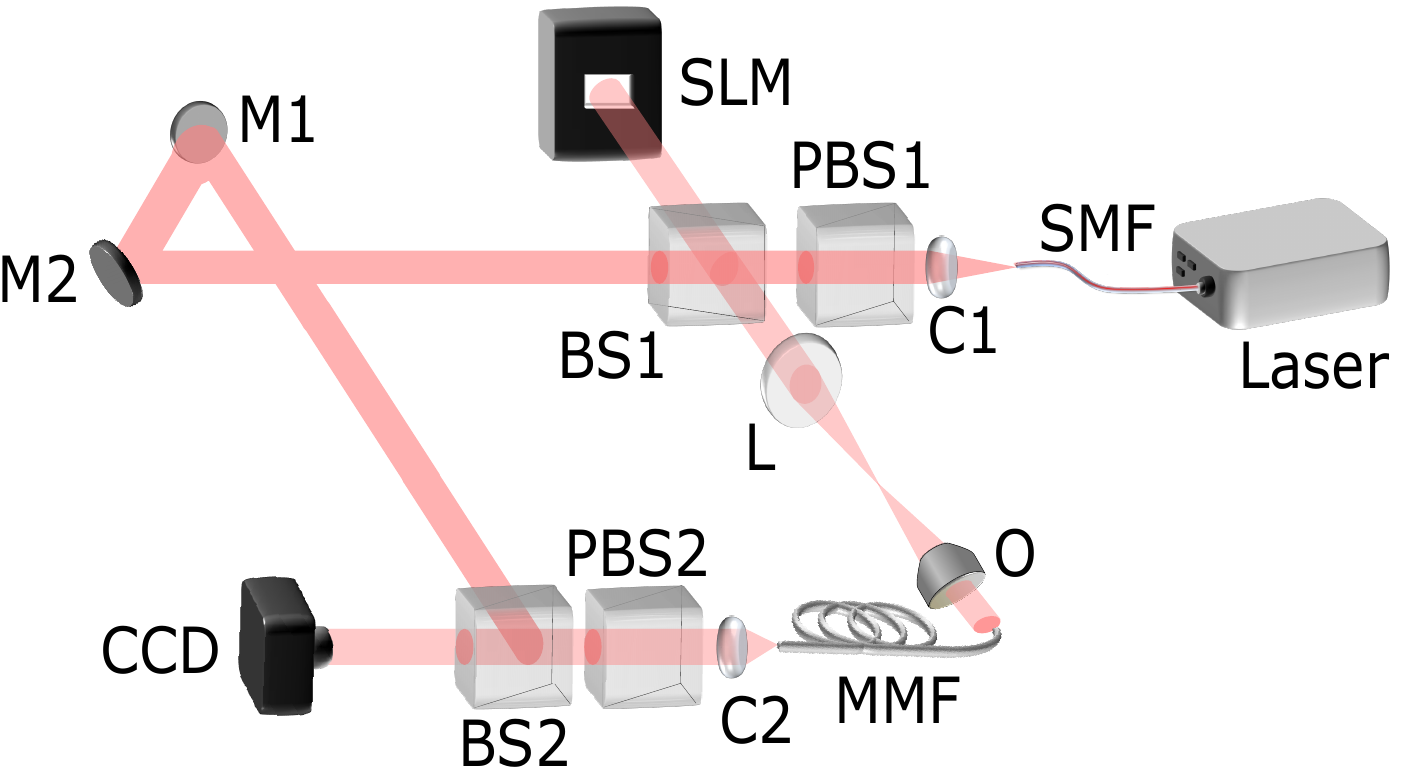}\caption{Experimental setup for measuring the field transmission matrix of a MMF. The continuous-wave output from a tunable laser source  (Agilent 81940A) at wavelength $\sim 1550\ \text{nm}$ is collimated (C1) and linearly polarized (PBS1). The beam is split into two arms by a beam splitter (BS1). In the fiber arm, light is modulated by the SLM in the reflection mode and then coupled to the MMF by a tube lens (L) and an objective (O). The output light from the MMF is collimated (C2) and linearly polarized (PBS2), before combining with the beam from the reference arm at a second beamsplitter (BS2). To match the optical path-length in the two arms, two mirrors (M1, M2) are inserted to the reference arm to adjust the path-length. BS2 is tilted to produce interference fringes of the two beams, which are recorded in the far field by a CCD camera. \label{fig:Setup}}
\end{figure}

To construct the Wigner-Smith time-delay matrix, we measured the field transmission matrices of a MMF at multiple wavelengths. Figure~\ref{fig:Setup} is a schematic of an interferometer setup.  A spatial light modulator (SLM) in the fiber arm prepares the phase front of the light field, which is then imaged to the front facet of a MMF. The output from the fiber combines with the reference beam and forms interference fringes. From the interferogram, we extract the spatial distribution of the transmitted field through the fiber. The measured intensity is $I= |E_r|^{2} + |E_s|^{2} + E_r^{*} E_s e^{ikr \sin{\theta}} + E_r E_s^{*} e^{-ikr \sin{\theta}}$, where $E_r$ and $E_s$ are the electric fields of the reference arm and the fiber arm, respectively, and $\theta$ is the tilt angle between them. The first two terms represent the dc components, and the last two terms are modulated at the spatial frequency $\pm k \sin \theta$. These terms can be separated in the Fourier domain, namely, by performing the spatial Fourier transform. By applying a Hilbert filter, we select only the third term that has the positive spatial frequency, then remove the factor $e^{ikr \sin{\theta}}$ before applying an inverse Fourier transform to obtain the amplitude and phase of $E_s$. 

\begin{figure}
\centering
\includegraphics[scale=0.28]{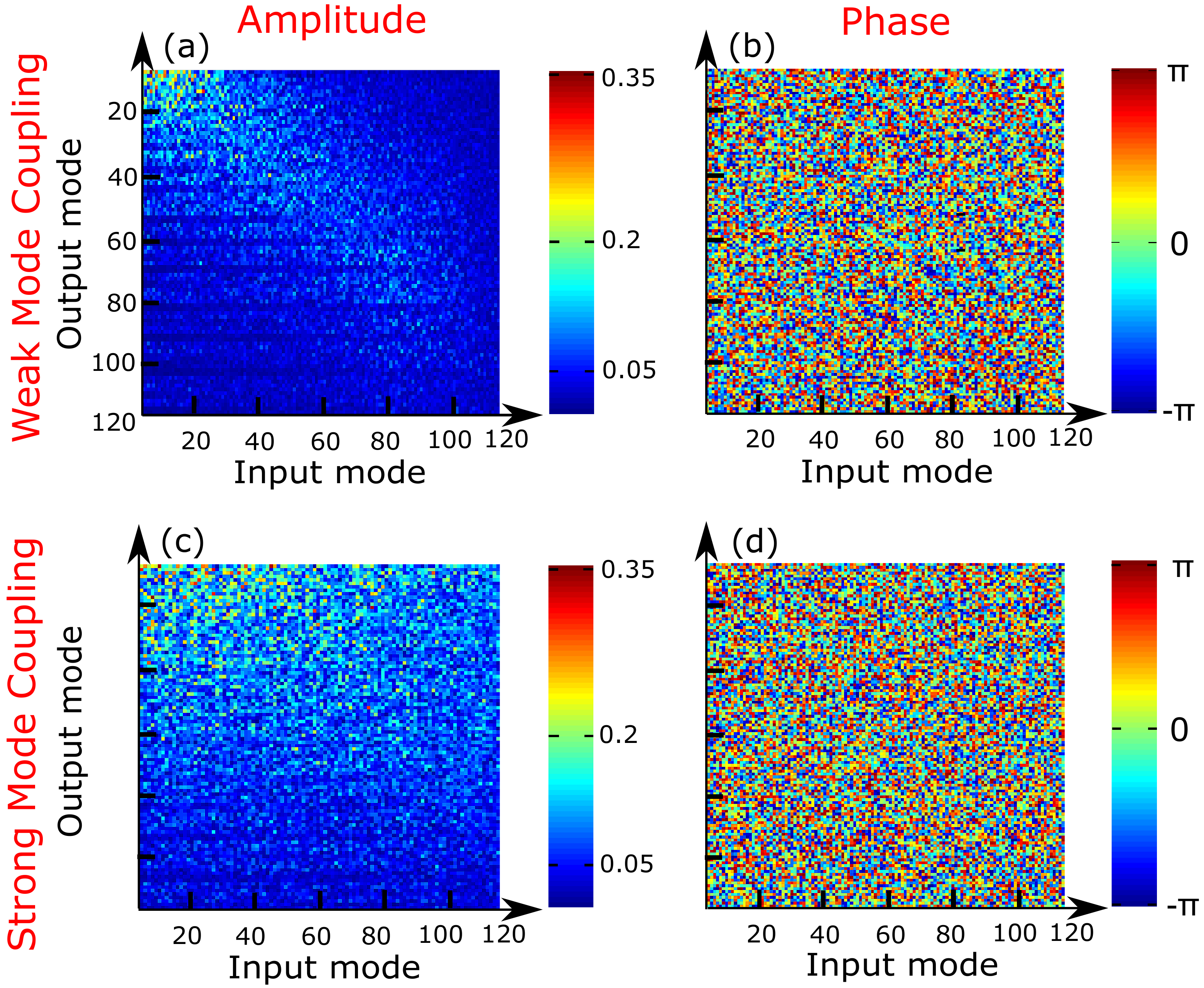}\caption{Field transmission matrix of a MMF with weak (a,b) and strong mode coupling (c,d). 
	The one-meter-long step-index fiber has a 50 $\mu$m core and a numerical aperture of 0.22.
	There are about 120 guided modes for one polarization, which are labeled by the propagation constant (from large to small). 
    Amplitude (a,c) and phase (b,d) of the measured transmission matrix at $\lambda$ = 1550 nm ($\omega$ = 194 THz) for one linear polarization. 
    The transmission matrix is nearly diagonal in (a), indicating weak coupling among modes of similar propagation constants. In (c) all modes are coupled, although higher-order modes have more attenuation.
\label{fig:Transmission matrices}}
\end{figure}

The transmission matrix is measured in momentum space. The SLM scans the incident angle of light onto the fiber facet, and the transmitted light is measured in the far field of the distal tip. We apply stress to the fiber with clamps to enhance the mode coupling. By adjusting the stress, we can tune the coupling strength. To evaluate the strength of mode coupling in the fiber, the transmission matrix is transformed to the mode basis by decomposing the input and output fields by LP modes, which are simply referred to as modes below. Figure \ref{fig:Transmission matrices} shows the amplitude and phase of the measured transmission matrices of the MMF. Without external stress, the field transmission matrix is nearly diagonal [Fig.~\ref{fig:Transmission matrices}(a)]. The small off-diagonal terms result from weak mode coupling due to inherent imperfection and macro-bending of the fiber. With an increase in the stress applied to the fiber, the off-diagonal terms grow and eventually become comparable to the diagonal terms, as shown in Fig.~\ref{fig:Transmission matrices}(c). Hence, in the weak coupling regime only modes with similar propagation constants are coupled. However, in the strong coupling regime, light diffuses to all modes regardless of which mode it is injected. Greater loss results in a lower amplitude of higher order modes at the output. However, if higher order modes are launched into the fiber, they can be scattered to lower order modes which experience less attenuation and dominate the output fields. Consequently, the transmission matrix presents a stronger decay for the output modes of high-order than the input ones. The phases of the transmission matrix elements are randomly distributed for $0$ and $2 \pi$, reflecting the random nature of the mode coupling in the MMF. 

\begin{figure}
\centering
	\includegraphics[scale=0.4]{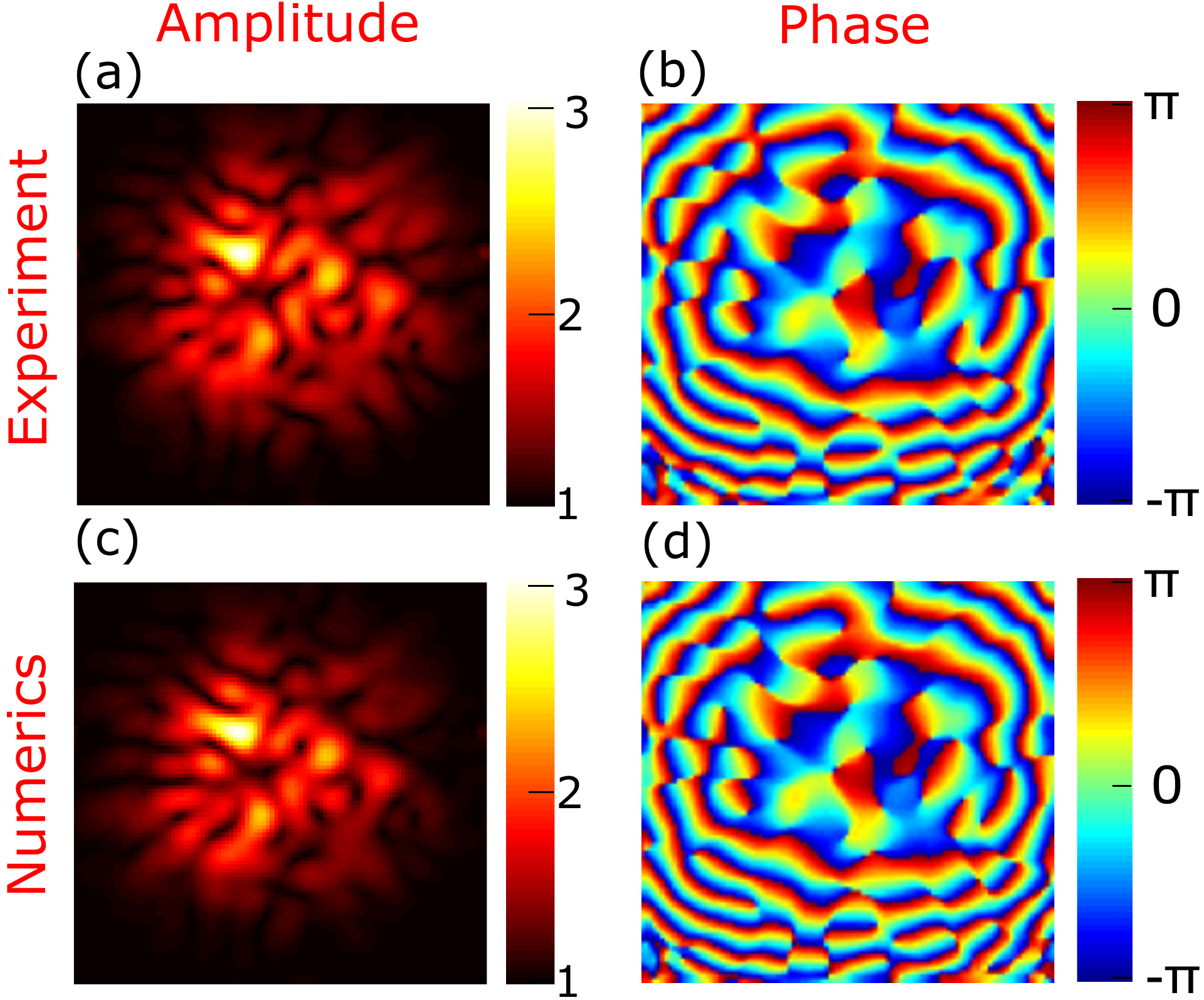}\caption{Experimental realization of PMs at $\lambda = 1550$ nm. 
		Experimentally measured (a,b) and numerically calculated (c,d) amplitude and phase of the output field of a PM in the same fiber as in Fig.~\ref{fig:Transmission matrices}. The agreement confirms the accuracy of the transmission matrix measurement. 
		\label{fig:PM}}
\end{figure}

After measuring the transmission matrices at multiple wavelengths, we construct the group-delay matrix $G \equiv -iT^{-1} {dT}/{d\omega}$. An eigenvector of $G$ gives the input field for a PM. We generate the input waveform of the principal mode by the SLM and launch it to the fiber. Since the SLM is limited to phase-only modulation, a complex-to-phase coding technique is used to convert the computer-generated phase-only hologram to a complex function with amplitude and phase modulation \cite{ArrizonJOSAA07}. Figure~\ref{fig:PM}(a,b) depict the measured amplitude and phase of the output field pattern $\Psi$ for a PM. For comparison, we also calculate the output field $\Psi'$ from the input field of the same PM using the measured transmission matrix [Figure~\ref{fig:PM}(c,d)]. To quantify their difference, we compute $\int|\Psi-\Psi'|^{2}d{\bf r}$, with $\int|\Psi|^{2}d{\bf r}=1$ and $\int|\Psi'|^{2}d{\bf r}=1$. The difference is $3.8 \%$, confirming the accuracy of our experimental measurement.

We note that the transmission matrix is measured for one linear polarization of input and output light only. Since the polarization is scrambled in the MMF, some of the input light is converted to the other polarization and thus is not measured at the output. The transmission matrix is non-unitary even without intrinsic loss, and it is part of the full transmission matrix for both polarizations. Nevertheless, we can still obtain the group-delay matrix for one polarization from the partial transmission matrix. Its eigenstate gives the linearly-polarized input waveform that generates an output field whose one polarization component has a frequency-invariant spatial profile. Below we study the characteristics of such polarized PMs, which are simply referred to as PMs. 

\section{Principal modes in weak and strong mode coupling regimes}
\begin{figure}
\centering
\includegraphics[scale=0.3]{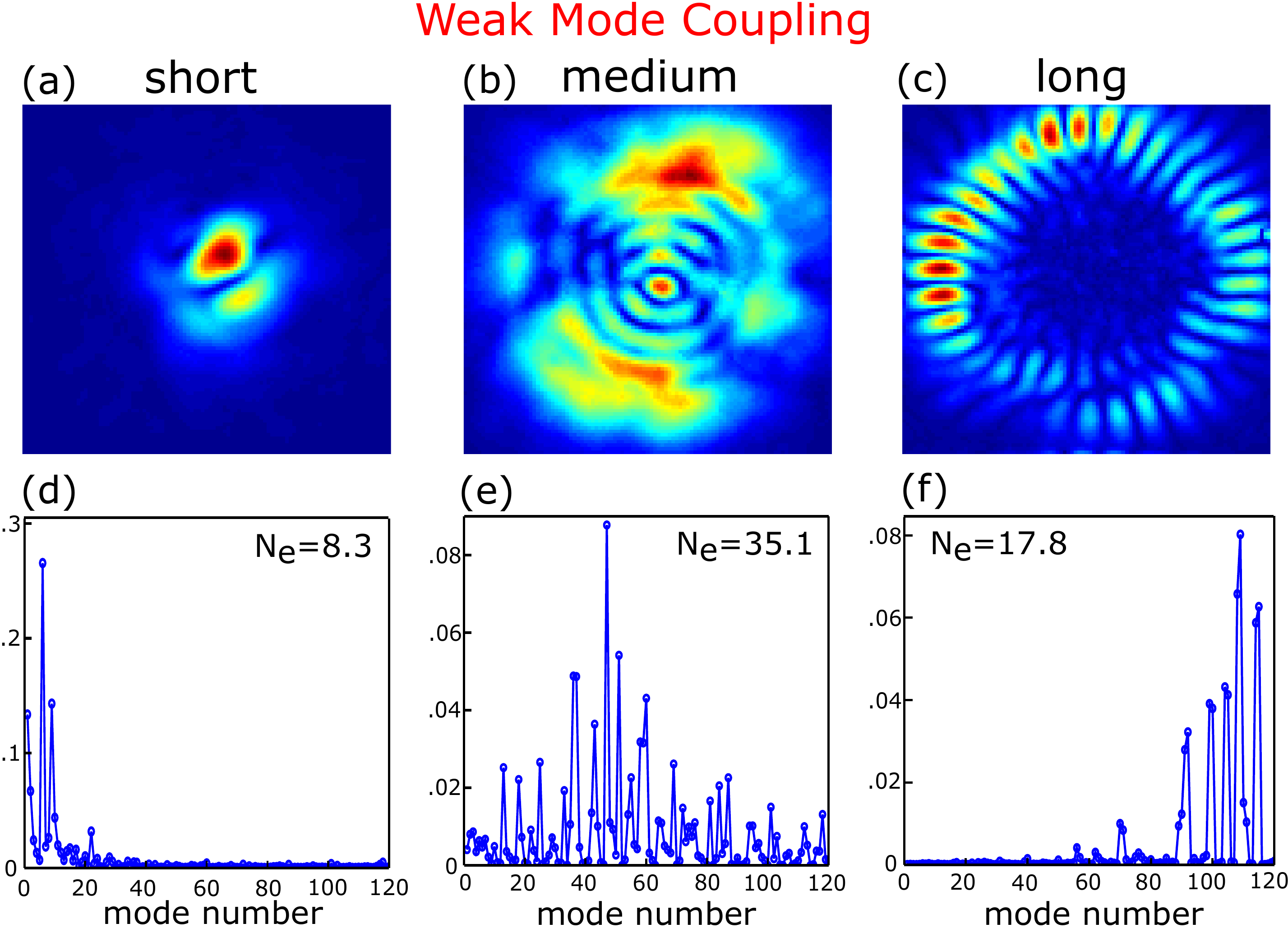}
\caption{PMs in the weak mode coupling regime. (a-c) Spatial distribution of the far-field amplitude for three PMs in the weak mode coupling regime with short (a), medium (b) and long (c) delay time. (d-f) Decomposition of output fields in (a-c) by the LP modes. The PMs with short/medium/long delay time are composed mostly of low/medium/high-order LP modes. $N_e$ is the mode participation number.
\label{fig: principal mode weak coupling}}
\end{figure}

We now experimentally investigate the differences in PMs of the MMF with weak and strong mode coupling. Figure~\ref{fig: principal mode weak coupling}(a-c) shows the far-field patterns of three PMs in the weak mode coupling regime with short, medium and long delay times. The PM with short delay time has small transverse momentum, similar to the low-order modes [Fig.~\ref{fig: principal mode weak coupling}(a)]. With increasing delay time, the PM acquires larger transverse momentum [Fig.~\ref{fig: principal mode weak coupling}(b)]. The far-field pattern of the PM with long delay time consists of large transverse momentum, like the high-order modes [Fig.~\ref{fig: principal mode weak coupling}(c)]. We decompose the output field pattern by the LP modes, and the coefficients are given in Fig.~\ref{fig: principal mode weak coupling}(d-f). The PM with short/medium/long delay time is composed mostly of low/medium/high-order modes. Hence, in the weak mode coupling regime, each PM contains only a few modes with similar propagation constants.

\begin{figure}
\centering
\includegraphics[scale=0.3]{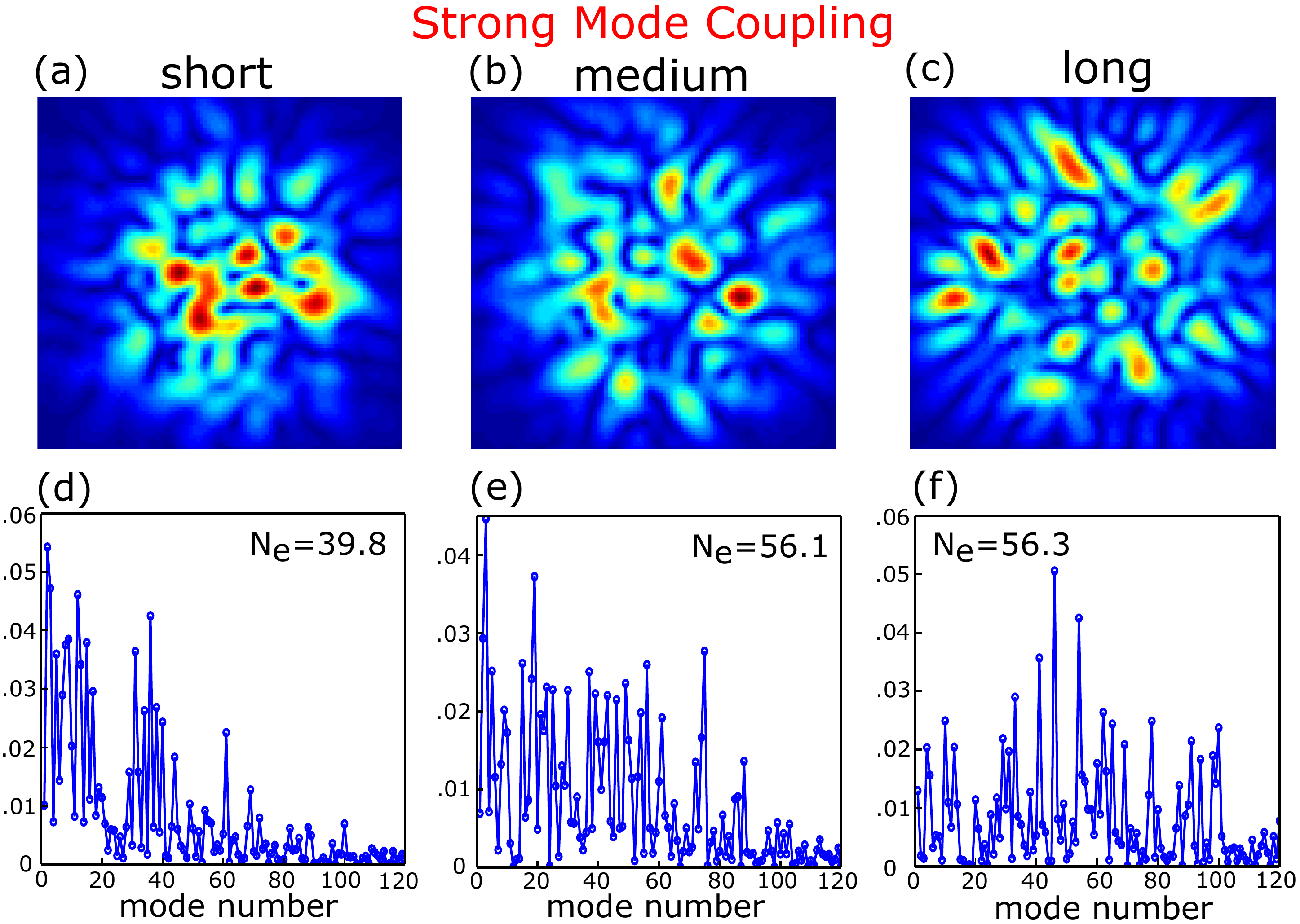}
\caption{PMs in the strong mode coupling regime. (a-c) Spatial distribution of the far-field amplitude for three PMs in the strong mode coupling regime with short (a), medium (b) and long (c) delay time. (d-f) Modal decomposition of output fields in (a-c), revealing the PM is a superposition of many LP modes. $N_e$ is the mode partition number.
\label{fig:principal mode strong coupling}}
\end{figure}

Figure~\ref{fig:principal mode strong coupling}(a-c) plots the spatial distribution of the output field amplitude for three PMs with short, medium and long delay time in the case of strong mode coupling. The far-field patterns contain many transverse momentum components and do not resemble any modes of the perfect fiber. The modal decomposition verifies that these PMs are a superposition of many LP modes [Fig.~\ref{fig:principal mode strong coupling}(d-f)]. Since higher-order modes experience more attenuation, their contributions to PMs, especially to the ones with shorter delay times, are reduced. To be more quantitative, we define the mode participation number as $N_e \equiv (\sum_n |\alpha_n|^2)^2/(\sum_n |\alpha_n|^4)$, where $\alpha_n$ is the decomposition coefficient for the $n$-th mode.  As noted in Figs.~\ref{fig: principal mode weak coupling} and \ref{fig:principal mode strong coupling}, the values of $N_e$ for the PMs in the weak mode coupling regime are significantly smaller than those in the strong mode coupling regime. 

Next we compare the spectral properties of PMs in the weak and strong mode coupling regimes with each other. For this purpose we scan the frequency $\omega$ while keeping the input field profile to that of a PM at a given frequency $\omega_0$. The output field pattern is measured at each frequency and correlated to that at $\omega_0$. We compute the spectral correlation function $C(\Delta\omega)=|\Psi(\omega_{0})\cdot\Psi(\omega_{0}+\Delta\omega)^{*}|$, where $\Psi(\omega)$ is a vector representing the output fields in all spatial channels, and its magnitude is normalized to one at each frequency. Figure~\ref{fig:correlation} (a,b) plots $C(\Delta\omega)$ for three PMs with short, medium and long delay times in weak and strong mode coupling regimes. For comparison, $C(\Delta\omega)$ for a random superposition of modes at the input is also shown.  The small revival of the correlation for the random input in the weak mode coupling regime is due to the spectral beating between different modes. One can clearly observe that the PMs decorrelate much more slowly with frequency detuning than the random input, and that they exhibit a plateau at $\Delta \omega = 0$. Moreover, the PM with short delay time decorrelates more slowly than the one with long delay time in both weak and strong mode coupling regimes. 

\section{PM bandwidth}
\begin{figure}
\centering
\includegraphics[scale=0.32]{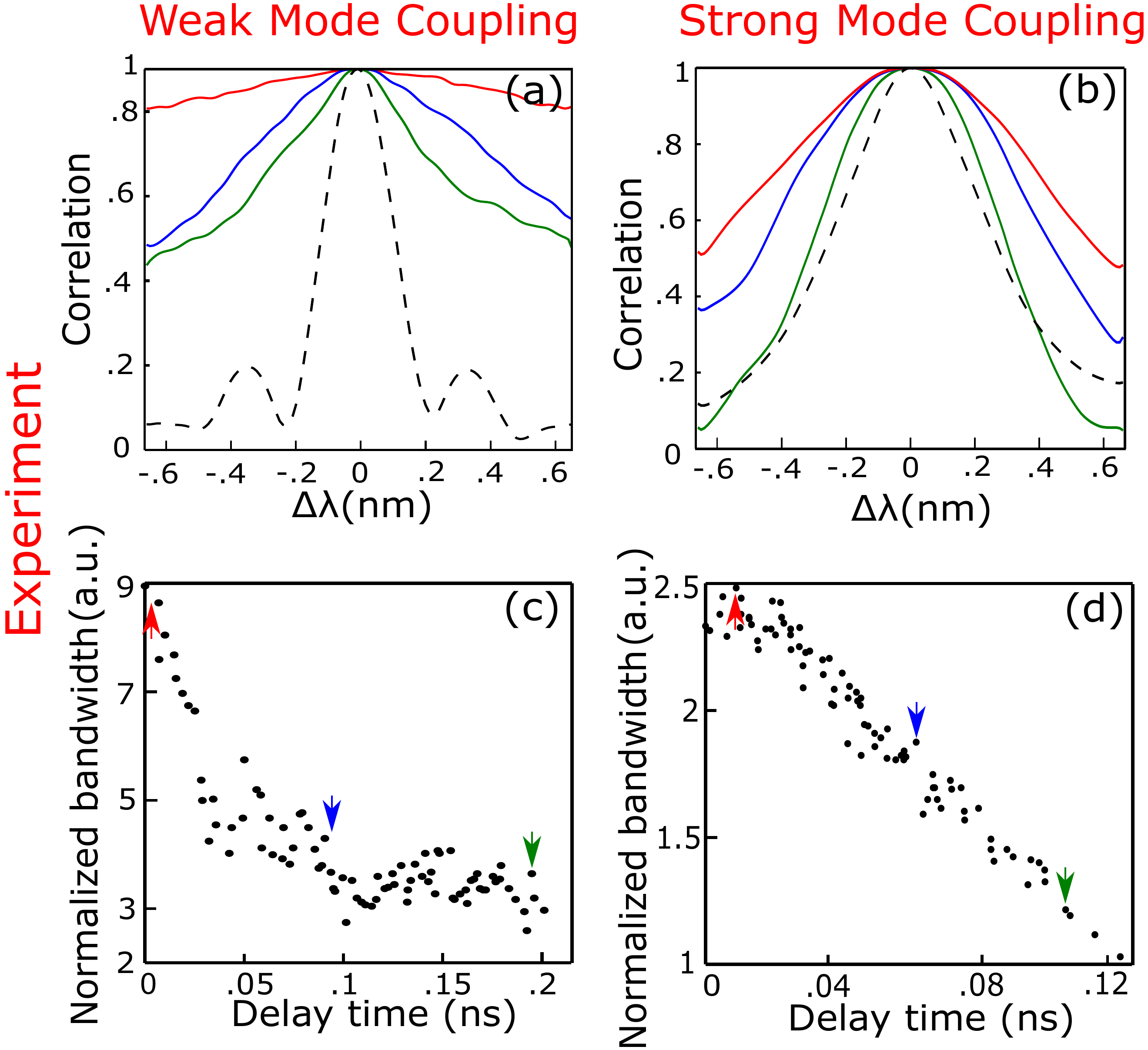}\caption{Spectral decorrelation of PMs. 
(a,b) Spectral correlation function $C(\Delta \omega)$ of the output field pattern, measured experimentally for three PMs with short delay time (red line), medium delay time (blue line) and long delay time (green  line) in the MMF with weak (a) and strong (b) mode coupling. For comparison, $C(\Delta \omega)$ for a random input is also shown (black dashed curve). $C(\Delta \omega)$ is normalized to one at $\Delta \omega=0$. The output field pattern for the PM with short delay time decorrelates more slowly with frequency than that with long delay time. 
(c,d) Normalized spectral correlation width of PMs vs. delay times in weak (c) and strong (d) mode coupling regime. The red, blue and green arrows indicate the PMs of which the three spectral correlation curves plotted in (a) and (b).
\label{fig:correlation} }
\end{figure}

To be more quantitative, we define the PM bandwidth $\Delta \omega_c$ as the frequency range over which $|C| \geq 0.9 |C(0)|$. Since the spectral decorrelation of the output pattern for any input waveform depends on fiber properties, such as the fiber length and numerical aperture, the PM bandwidth is normalized by the average correlation width of random inputs.  Figure~\ref{fig:correlation} (c,d) plot the normalized $\Delta \omega_c$ for all PMs versus their delay times. In the weak mode coupling regime the PM bandwidth first drops sharply with increasing delay time, then levels off. In the strong mode coupling regime,  $\Delta \omega_c$ remains nearly constant at short delay time, and starts decreasing as the delay time becomes larger. The normalized bandwidths of PMs in the weak mode coupling regime are larger than those in the strong mode coupling regime, indicating the PMs in the presence of weak mode coupling decorrelate slower than those with strong mode coupling. 

To understand what determines the bandwidths of PMs in the MMFs with weak and strong mode coupling, we perform numerical simulations using the concatenated fiber model \cite{HoJLT11}. In particular, we consider a one-meter-long step-index fiber with 50 $\mu$m core and 0.22 numerical aperture. The fiber is divided into 20 short segments; light propagates in each segment as in a perfect fiber without mode coupling. Between adjacent segments, the guided modes are randomly coupled. The scattering in the mode space is simulated by a unitary random matrix, which is given by $A=\exp[iH]$, where $H$ is a random Hermitian matrix. We construct $H=G \cdot (R+R^{\dagger})$, in which $R$ is a complex random matrix whose elements are taken from the normal distribution, and $G$ is a real matrix imposing a Gaussian  envelope function on the matrix elements along the off-diagonal direction. Specifically, the magnitude of the matrix elements decays away from the diagonal, and the decay rate, i.e., the width of the Gaussian envelope function, depends on the degree of mode coupling. The faster the decay, the narrower the envelope function and the weaker the mode coupling. Therefore, by varying the width of the Gaussian envelope function, we can tune the scattering strength in mode space. 

To quantify the amount of scattering in mode space,  we calculate the effective transport mean free path $\ell$, which is given by the propagation distance beyond which the spatial field profile becomes uncorrelated \cite{HoJLT14}. In the concatenated fiber model, the transport mean free path is obtained numerically by launching light into a single mode and computing the number of segments light propagates until all modes are equally populated. The coupling strength is described by the ratio of the fiber length $L$ to the effective transport mean free path $\ell$.

\begin{figure*}
\includegraphics[scale=0.4]{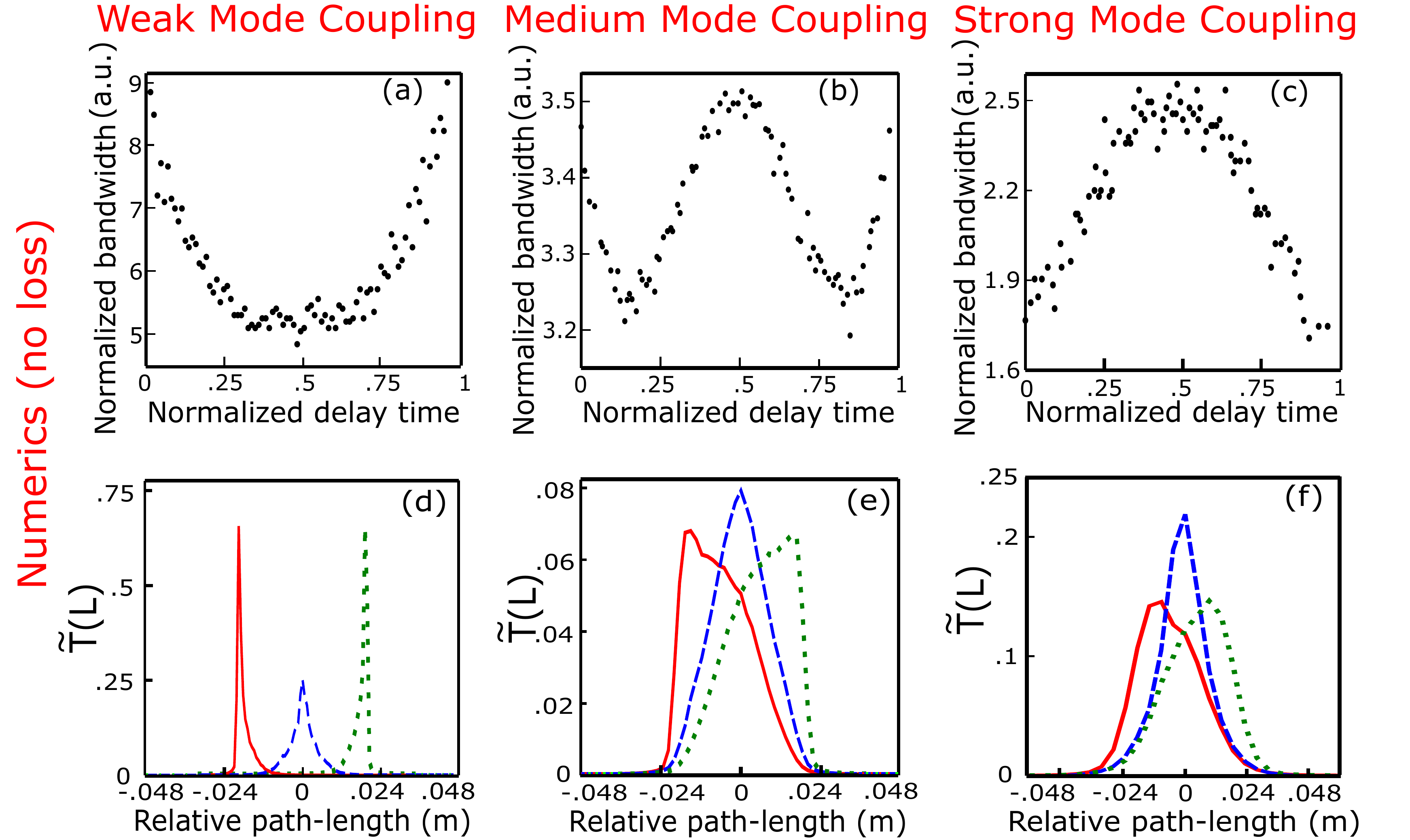}\caption{Calculated PM bandwidths (upper row) and corresponding path-length distributions (lower row). 
	The MMF is a step-index fiber with 50 $\mu$m core and 0.22 numerical aperture. The mode coupling strength ($L/\ell$) is 0.2 in (a,d), 1.0 in (b,e), and 10 in (c,f).   
(a,b,c) PM bandwidths vs. delay times. The bandwidth is normalized to the average width of random inputs.  The shortest delay time is set to be 0 and the difference between the shortest and longest delay time is normalized to 1. 
(d,e,f) The intensity distributions over the relative path-length of the PMs in (a,c,e) with the delay time = 0  (red solid line), 0.5 (blue dashed line) and 1 (black dotted line). 
In the weak mode coupling regime (a), the PM bandwidth is maximized at the shortest and longest delay time. In the strong mode coupling regime (c), the PM bandwidth is the largest at the medium delay time. (b) shows the transition between the two regimes. 
\label{fig:Bandwidth-of-principal_no_loss}}
\end{figure*}

First we ignore the fiber loss and calculate the normalized bandwidths of PMs in the MMF with different degrees of mode coupling. In the weak coupling limit ($L/ \ell \ll 1$), the PM bandwidth has two maxima at the shortest delay time and the longest delay time [Fig.~\ref{fig:Bandwidth-of-principal_no_loss}(a)]. As the mode coupling ($L/ \ell$) increases, the normalized bandwidths of all the PMs are reduced. However, the decrease at the medium delay time is slower than that at short and long delay times. Consequently, a new maximum arises at the medium delay time when ($L/ \ell \simeq 1$) [Fig.~\ref{fig:Bandwidth-of-principal_no_loss}(b)]. With a further increase of mode coupling, the two local maxima at the shortest and longest delay times disappear entirely [Fig.~\ref{fig:Bandwidth-of-principal_no_loss}(c)]. Thus the variation of the bandwidth with the delay time in the strong mode coupling regime  ($L/ \ell \gg 1$) is just opposite to that in the weak mode coupling regime. 

To interpret these results, we resort to an intuitive picture of optical paths in the MMF. A MMF supports many propagating modes, each having a different propagation constant. From the geometrical-optics point of view, various rays propagate down the fiber at different angles relative to the axis of the fiber, and thus travel different distances and experience different phase delays. Inherent imperfections and external perturbations result in light hopping among the trajectories with different angles and lengths. Hence, light can take many paths of different lengths to transmit through the fiber. The sum of waves following different paths gives the output field. Formally, this fact can be expressed by writing the transmission amplitude $t_{nm}(\Delta\omega)$ from an incoming mode $m$ and an outgoing mode $n$ at frequency $\Delta\omega$ (central frequency is set to zero) through a sum over infinitely many paths $q$, each of which contributes with an amplitude $A_q$ and with a phase that depends on the path length $L_q$ in the following way: $t_{nm}(\Delta\omega) = \sum_q A_q \exp(i\Delta\omega L_q/c)$ \cite{JalabertPRL90}. This relation follows directly from the Feynman path integral formulation of the Green's function, for which several semiclassical approximations have been worked out (see \cite{Brack97} for an overview). The interesting insight that we now deduce from this path picture is that in the weak guiding approximation one can deduce the path spectrum $\tilde{t}_{nm}(L)$ contributing to the transmission amplitude $t_{nm}(\Delta\omega)$ by a simple Fourier transform $\tilde{t}_{nm}(L)=\int_{k_{\rm min}}^{k_{\rm max}}dk\,t_{nm} (k)\exp(-ikL)$ \cite{RotterPRB00}, where we define $k = \Delta\omega/c$. Correspondingly, the power spectrum of the total transmission through the fiber is given as $\tilde{T}(L)=\sum_{n,m}|\tilde{t}_{nm}(L)|^2$.

The width of the intensity distribution over the path-length spectrum determines how fast the output field decorrelates with frequency. The narrower the distribution, the weaker the dephasing among different paths, and the slower the decorrelation. We calculate $\tilde T(L)$ for the PMs in different mode coupling regimes. Figure~\ref{fig:Bandwidth-of-principal_no_loss}(d,e,f) presents the results for three PMs with the shortest, intermediate, and longest delay times. In the case of weak mode coupling, the intensity distribution over the path-length is narrow [Fig.~\ref{fig:Bandwidth-of-principal_no_loss}(d)] because each PM contains only a few modes with similar propagation constants. For example, the PM with short delay time consists of a few low-order modes. The adjacent modes that these low-order modes can couple to are higher order modes with smaller propagation constants. However, the PM with intermediate delay time is composed of modes with medium propagation constants, which are surrounded by both lower and higher order modes to which they can couple to. Since the propagation constants of modes in a step-index fiber are almost equally spaced, the constituent modes for an intermediate PM have more neighboring modes to couple to, and the intensity distribution over the path-length is wider than that for the fast PM. Consequently, the fast PM has a broader bandwidth than the intermediate PM. The same argument applies to the slow PM that has a long delay time. Therefore, the fastest and slowest PMs have the maximum bandwidth.

As the mode coupling strength increases gradually, the intensity distribution over the path-length is broadened [Fig.~\ref{fig:Bandwidth-of-principal_no_loss}(e)], and the bandwidth of PMs is reduced. Eventually all modes are coupled, and the transition from single scattering to multiple scattering occurs in mode space. In the regime of multiple scattering, wave interference becomes significant. Since light can follow many possible trajectories of the same length from the input to the output of the fiber, the interference of the fields from these paths determines the intensity distribution over the path-length. In Fig.~\ref{fig:Bandwidth-of-principal_no_loss}(f), the fast PM has intensity concentrated on shorter paths, as the destructive interference of different trajectories with the same length makes $\tilde T(L)$ is mere pronounced for longer path-length. The opposite happens to the slow PM. Quite remarkably, Such interference effects are completely determined by the input wavefront. PMs with intermediate delay times suppress both short and long paths by destructive interference. In the absence of mode-dependent loss, the central-limit-theorem dictates that the density of path-lengths has a Gaussian distribution that is peaked at the medium delay time \cite{XiongPRL16}. Thus the intermediate PMs, whose delay times coincide with or are close to the medium path-length of maximal density, only need to suppress a small number of trajectories of short or long path-lengths via interference. By contrast, the PMs with short delay times require destructive interference of both medium and long paths. Since there are more trajectories with medium path-length, it is more difficult to suppress them via interference, as evident from the shoulder at medium path-length for the fast PM in Fig.~\ref{fig:Bandwidth-of-principal_no_loss}(e). Hence, the fast PMs have broader path-length distributions and narrower bandwidths than the medium PMs. The same explanation applies to the bandwidths of the slow PMs.

\begin{figure}
\centering
\includegraphics[scale=0.42]{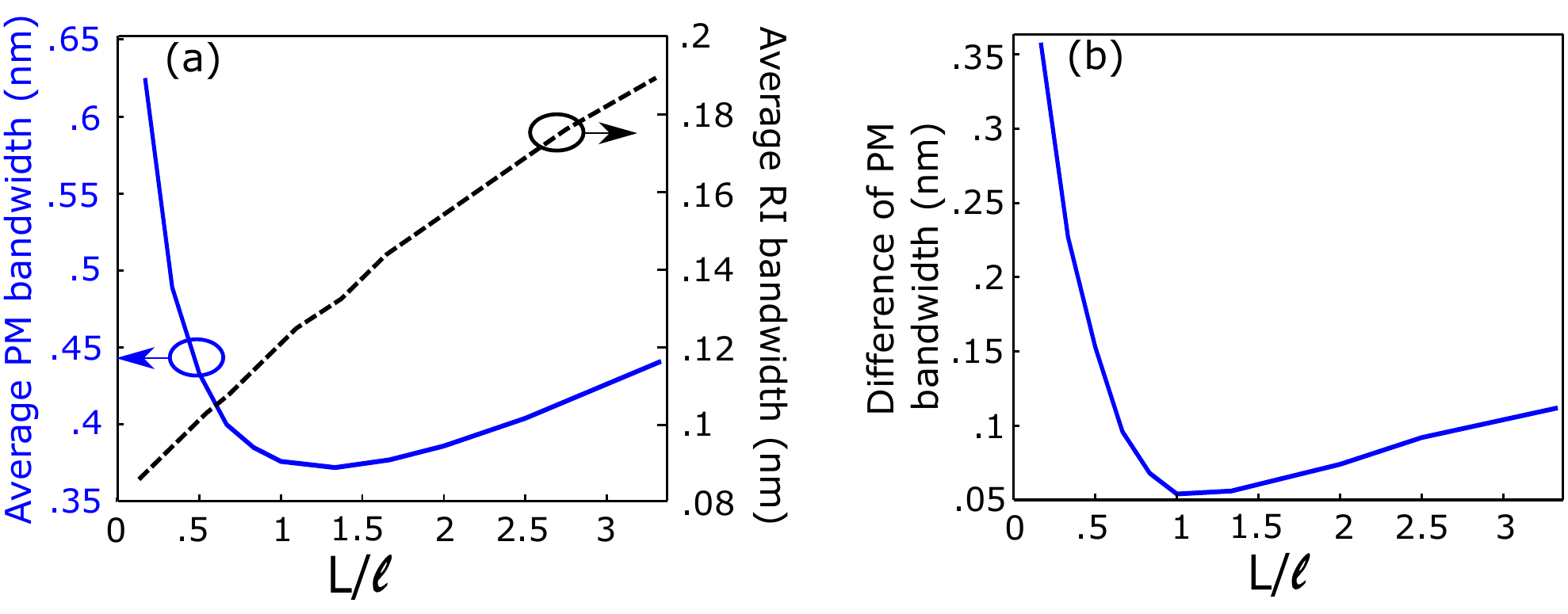}\caption{Evolution of PM bandwidth with mode coupling strength.
(a) The average bandwidth of all PMs (blue solid curve) first decreases with $L/\ell$, reaches the minimum at $L/\ell \simeq 1$, and then increases. For comparison (black dashed curve), the average bandwidth of random inputs increases monotonically with $L/\ell$.
(b) The difference between the maximum bandwidth and the minimum bandwidth exhibits a similar trend as the average bandwidth. 
\label{fig:absolute_bandwidth}}
\end{figure}

We further analyze the transition from weak to strong mode coupling. As $L/ \ell $ increases, the average bandwidth of random input fields increases monotonically, as shown by the black dashed line in Fig.~\ref{fig:absolute_bandwidth}(a). For PMs, the average bandwidth first decreases rapidly, then goes through a turning point at $L/ \ell \simeq 1$, and starts increasing again [Fig.~\ref{fig:absolute_bandwidth}(a), blue solid curve]. In the single scattering regime $L / \ell < 1$, the input light spreads further in mode space as the scattering strength increases, and each PM consists of more LP modes. In particular, the number of LP modes in the PM with short or long delay time grows faster and approaches that with medium delay time. Consequently, the path-length distributions broaden more quickly and the bandwidths decrease more rapidly  for the slow and fast PMs, leading to the reduction of the two local maxima at the shortest and longest delay times [Fig.~\ref{fig:Bandwidth-of-principal_no_loss}(a,b)].

Once $L / \ell$ exceeds 1, the light is coupled back and forth among the modes, and the interference effects arise. In particular, the multi-path interference narrows the intensity distribution over the path-length spectrum. Stronger scattering enhances the interference effects, leading to an increase of the average bandwith of PMs [Fig.~\ref{fig:absolute_bandwidth}(a), blue solid curve]. Since the multi-path interference effect is more efficient in narrowing the path-length distribution for a PM with intermediate delay time, its bandwidth is broader than that of a PM with short or long delay time. Hence, a local maximum in the bandwidth arises at the medium delay time, as seen in Fig.~\ref{fig:Bandwidth-of-principal_no_loss}(b,c). 

In Fig.~\ref{fig:absolute_bandwidth}(a), the average PM bandwidth exhibits a minimum at the transition point ($L / \ell \simeq 1$) from single scattering to multiple scattering in the mode space. At this point, light is spread over all LP modes, yet the multipath interference effect is not yet strong enough to enhance the PM bandwidth. To investigate the fluctuation of PM bandwdiths, we also calculate the difference between the largest and smallest bandwidth of PMs, which exhibits a trend similar to the average bandwidth as seen in Fig.~\ref{fig:absolute_bandwidth} (b). In the weak mode coupling regime, the difference is large but it declines dramatically with the coupling strength. When the system gradually transits to the strong mode coupling regime, the difference increases slightly, but still remains at a small value. 

\section{Effect of mode-dependent loss}
\begin{figure}
\centering
\includegraphics[scale=0.34]{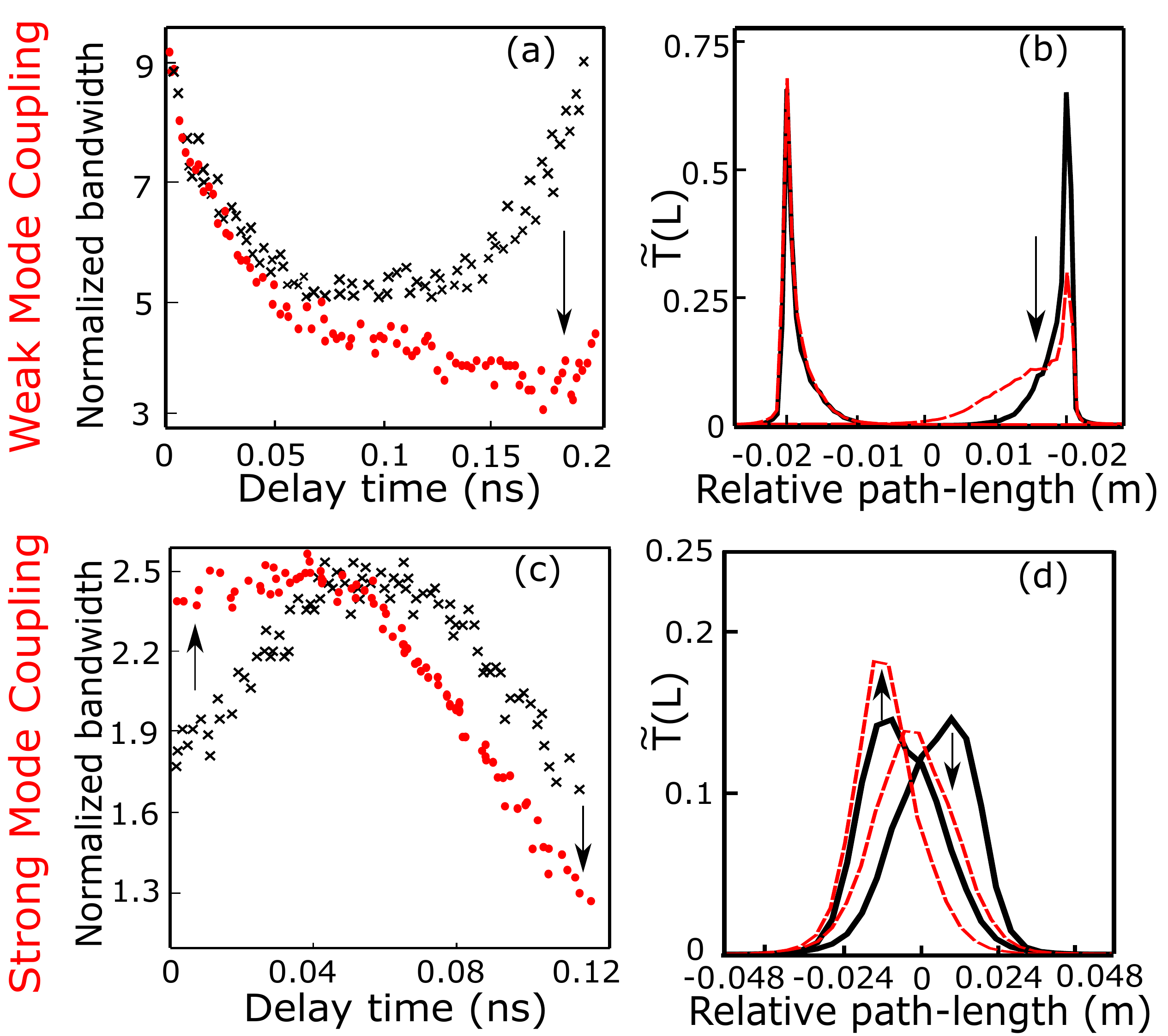}\caption{Effects of MDL on PM bandwidths and path-length distributions. (a) Normalized bandwidths of PMs in the weak (a) and strong (c) mode coupling regimes with (red dots) or without (black crosses) MDL. (b,d) Calculated intensity distributions over the path-length of PMs in (a,c) with delay time = 0, 0.2 ns in (b) and 0, 0.12 ns in (d) with (red dashed) or without (black solid) MDL. In the weak coupling regime, the MDL reduces significantly the bandwidth of slow PMs (a) by broadening their path-length distributions (b). In the strong coupling regime, the MDL enhances the bandwidth of fast PMs (c) by narrowing their path-length distribution (d).    \label{fig:Bandwidths with loss}}
\end{figure}

The numerical study in the last section assumes no loss in the fiber. However, loss is common in a MMF, and it is usually greater for higher-order modes. In this section, we investigate the effects of mode-dependent loss (MDL) on PMs. In the concatenated fiber model, we introduce a uniform absorption coefficient to each segment of the fiber. Higher order modes that have longer transit time thus experience more loss. 

We compare the PM bandwidth with MDL to that without MDL in Fig.~\ref{fig:Bandwidths with loss}. In the weak mode coupling regime, MDL significantly reduces the bandwidth of PMs with long delay times, as indicated by the arrow in Fig.~\ref{fig:Bandwidths with loss}(a). In contrast, the bandwidth of PMs with short delay time are nearly unchanged by the MDL. This behavior can be explained by the change in the intensity distribution over the path-length $\tilde T(L)$. The slow PM is composed of long paths, and the stronger attenuation of the longer paths broadens the distribution, as shown in Fig.~\ref{fig:Bandwidths with loss}(b). Consequently, the bandwidth of the PM with long delay time is reduced. The fast PM, by contrast, consists of short paths, which experience little loss, thus $\tilde T(L)$ remains almost the same, and with it also the bandwidth of the PM. The longer the delay time, the stronger the effect of MDL, and the greater the reduction in the PM bandwidth. 

In the strong mode coupling regime, the MDL enhances the bandwidth of a PM with short delay time, while reducing the bandwidth of PM with long delay time [Fig.~\ref{fig:Bandwidths with loss}(c)]. Since the fast PM has a broader path-length distribution than that in the weak mode coupling regime, the MDL suppresses the longer paths and narrows the distribution that centers on the short path-length [Fig. \ref{fig:Bandwidths with loss}(d)]. In contrast, the path-length distribution for the slow PM, which centers on the long path-length, is broadened by the MDL, as the shorter paths experience less attenuation than the longer ones. The variations of the PM bandwidth with the delay time in both weak and strong coupling regimes agree qualitatively with the experimental results in Fig. \ref{fig:correlation}(b,d). We may thus conclude that MDL has a significant impact on the bandwidths of PMs and needs to be taken into account to understand the experimental data.  

\section{Conclusion}
We have performed experimental and numerical studies on the principal modes (PMs) in a multimode fiber, which are the eigenstates of the Wigner-Smith time-delay operator or the group delay operator. By applying external stress to the fiber and gradually adjusting the stress, we have realized the transition from weak to strong mode coupling. Such a transition is mapped to that from single scattering to multiple scattering in mode space. We experimentally demonstrate that PMs have distinct spatial and spectral characteristics in weak and strong mode coupling regimes. In the weak mode coupling regime, each PM is composed of a small number of fiber eigenmodes with similar propagation constants. In the strong mode coupling regime, however, a PM is formed by all modes. When there is no mode-dependent loss in the fiber, PMs with shorter or longer delay times have broader bandwidths in the weak mode coupling regime. The opposite is true for strong mode coupling where the bandwidth is maximal for PMs with medium delay times. By analyzing the path-length distributions, we discover two distinct mechanisms that determine the bandwidth of PMs in the weak and strong mode coupling regime. For weak mode coupling, fast or slow PMs spread less in mode space and experience weaker modal dispersion, thus having broader bandwidth than intermediate PMs. In the presence of MDL, the bandwdith for a slow PM is reduced significantly while that for a fast PM remains nearly unchanged. In the strong mode coupling regime, interference among numerous trajectories in the multimode fiber becomes significant, and the maximum bandwidth is reached for the PMs whose delay time corresponds to the maximum density of path-length. Without MDL, the density of path-length is peaked at intermediate lengths such that the PMs with medium delay time have the largest bandwdith. With MDL, the maximum density of path-length shifts to shorter paths, due to stronger attenuation of longer paths in the fiber. Consequently, MDL enhances the bandwidth of fast PMs while it reduces the bandwdith of slow PMs.

\section*{Funding}
This work is supported partly by the US National Science Foundation under the Grant No. ECCS-1509361 and by the US Office of Naval Research under the MURI grant No. N00014-13-1-0649. P.A. and S.R. acknowledge support by the Austrian Science Fund (FWF) through projects SFB NextLite (F49-P10) and project GePartWave (I1142). 

\section*{Acknowledgments} We acknowledge Chia Wei Hsu, Nicolas Fontaine, Tsampikos Kottos, and Boris Shapiro for simulating discussions. 

\email{hui.cao@yale.edu}

\end{document}